\documentclass[twocolumn,10pt,showkeys,preprintnumbers,amsmath,amssymb]{revtex4}
\usepackage{graphicx}
\usepackage{dcolumn}
\usepackage{bm}
\usepackage{amsmath}
\usepackage{enumerate}
\newcommand {\me} {e}
\newcommand {\mi} {i}
\newcommand {\dif} {d}
\newcommand {\mj} {j}
\newcommand {\epsln} {\epsilon}
\begin{document}

\title{Statistics of Impedance and Scattering Matrices \\of Chaotic   Microwave
Cavities with  Multiple Ports
}  
\author{Xing Zheng}
\author{Thomas M. Antonsen Jr. }
\altaffiliation[Also at ] {Department of Electrical and Computer
Engineering.}
\author{Edward Ott}
\altaffiliation[Also at ] {Department of Electrical and Computer
Engineering.}\affiliation{Department of Physics\\ and Institute
for Research in Electronics and Applied Physics,\\ University of
Maryland, College Park, MD, 20742}
\date{\today}

\begin{abstract}
The statistical model proposed in an accompanying paper is generalized to treat multiport
scattering problems. Attention is  first focused on two-port lossless systems  and the model is
shown to be consistent with Random Matrix Theory. The predictions are then tested by direct
numerical simulation for a specific two-port cavity. Formulae are derived for the average
transmission and reflection coefficients in terms of the port radiation impedance. The cases of
cavities with multiple ports, and with a single port and distributed losses are compared.
\end{abstract}


\maketitle                 


\section{\label{sec:level1}Introduction} 

In an accompanying paper \cite{paperpart1}, a general statistical
model is proposed to describe scattering of high frequency electromagnetic
waves
from irregular cavities (Here by high frequency we mean that the
wavelength is substantially less than an appropriate characteristic
length of the cavity.). This problem arises when one considers
the coupling of electromagnetic energy into and out of complicated
enclosures. The statistical approach is warranted when the exact
details of the configuration are unknown or are too complicated to
simulate accurately. The model proposed in Ref.~\cite{paperpart1}
gives a method for calculating the statistical distributions of
important quantities that depend only on a small number of
system-specific parameters. 
These system-specific parameters are the
average spacing between
resonant frequencies of the enclosure, the average quality factor,
and the properties of the coupling ports. It was shown
\cite{paperpart1} that the key property needed to characterize  the
coupling port is
its radiation impedance, that is, the impedance that would be seen
at the port if the other boundaries of the enclosure were
perfectly absorbing or were removed to infinity. This impedance can be
used to ``normalize"
the fluctuating impedance at the port of the actual enclosure.
With this normalization the fluctuating impedance has a universal (i.e.,
system-specific independent)
distribution.

The treatment in Ref.~\cite{paperpart1}  focused on the case of a
single port. The purpose of the present paper is to generalize these
results to the case of multiple ports. A multiple port description is
necessary if one is to describe the transmission of energy entering the
enclosure to some other point in the enclosure. We will show that the
multiple port case can be treated using the same approach as the single
port case. The main difference being that the multiple port description
involves the introduction of matrices describing the coupling to the
system.

Scattering can be described by either the impedance matrix, which
relates voltages and currents at the ports, or by the scattering
matrix, which relates the amplitudes of incoming and outgoing waves. In
our model we focus initially on the impedance matrix because its
properties can be related directly to the radiation impedances of the
ports. Then based on our understanding of the impedance matrix,
conclusions about the scattering matrix are drawn.

The statistical treatment of wave scattering in complicated
systems has been developed extensively in the physics community in
connection with nuclear scattering \cite{krieger67} and scattering
in mesoscopic systems such as quantum dots \cite{alhassid00} and
disordered conductors \cite{mello90}. Many of the concepts
developed in consideration of these problems have also been
applied to electromagnetic scattering in microwave cavities
\cite{doron90, anlage00, stockmann03, hemmady04, barthelemy04}.
The underlying approach is a description of the system based on
random matrix theory (RMT)\cite{seba96, theo4, mehta91}. Here, the
specific systems under consideration are replaced or modelled by
random matrices. These are matrices whose elements are independent
Gaussian random numbers drawn from specific ensemble
distributions. Two different ensembles that are considered are
relevant to our problem. The Gaussian Orthogonal Ensemble (GOE)
gives rise to a real symmetric matrix, where the variance of the
diagonal elements is twice that of the off-diagonal elements. This
applies to wave systems with time reversal symmetry (TRS), as
would apply in media with real symmetric permittivity and
permeability tensors. The Gaussian Unitary Ensemble (GUE) gives
rise to complex, self-adjoint matrices, where the off-diagonal
elements have independent real and imaginary parts.  This applies
to wave systems that have time reversal symmetry broken (TRSB).
For electromagnetic wave systems, this case is realized when a
nonreciprocal medium such as magnetized ferrite or a magnetized
cold plasma is present in the system.

Random matrix theory, in spite of its high level of abstraction,
has been remarkably successful in predicting the universal
statistical properties of wave systems. This includes the
description of the distribution of resonances in closed systems as
well as the properties of scattering from open systems. One issue,
that we address here is how to simultaneously account for the
universal properties as predicted by RMT and the system-specific
properties that depend on details of the coupling. In our approach
this connection is made through the radiation impedance of the
ports \cite{hemmady04} (This connection has also been made by Warne et al.
\cite{warne03}). An alternative approach \cite{mello85} in the
physics community, known as ``the Poisson Kernel", is based
directly on the scattering matrix. This approach describes the
distribution of fluctuations in the scattering matrix elements  in
terms of their average values. Recently, Kuhl et.
al.\cite{stockmann04} applied the Poisson Kernel approach to
analyse the statistics of measured values of the reflection
coefficeint for a microwave cavity. The measured results were
found to be in agreement with predictions based on the Poisson
kernel.

The organization of this paper is as follows. Section II
generalizes the description of the quasi-two dimensional cavity
introduced in Ref.~\cite{paperpart1} to describe the presence of
multiple ports. Our treatment will be sufficiently general so as
to treat the case of both isolated ports and non-isolated ports.
We then discuss the changes that must be made when non-reciprocal
elements such as a magnetized ferrite are present. In Sec. III we
examine the statistical properties of the proposed impedance
matrix, focusing on the two-port case. We also compare the model
predictions with the results of computational electromagnetic
calculations using HFSS. In Section IV we consider properties of
the scattering matrix when the port coupling is imperfect. We find
that the average value of the reflection coefficient is a function
of the radiation reflection coefficient. Also,  we consider the
case of multiple ports and verify the equivalence of distributed
and diffractive losses in the case of  large number of ports.
Section V contains a summary of our results. Except for Section
IV(B) all our considerations will be in the context of lossless
cavities.

\section{Generalization of the Model}
\subsection{The Impedance in the TRS Case}
Following Ref.\cite{paperpart1} we consider a quasi-two
dimensional cavity in which only the lowest order transverse
magnetic modes are excited. The fields in the cavity are
determined by the spatially dependent phasor amplitude of the
voltage $\hat{V}_T(x,y)$. The voltage is excited by currents
$\hat{I}_i$ drives at the various coupling ports,
\begin{equation}
(\nabla_{\perp}^{2}+k^{2})\hat V_T= -\mj
kh\eta_{0}\sum^M_{i=1}u_{i}\hat{I_{i}}. \label{eq:virelation2}
\end{equation}
Here $k=\omega /c$, $\eta_0=\sqrt{\mu_0/\epsln_0}$, $h$ is the
height of the cavity and an exponential time dependence
$\exp(j\omega t)$ has been assumed for all time dependent
quantities. Each of the $M$ ports is characterized by a profile
function $u_i$ centered at different locations and $\int \dif
x\dif y u_i=1$. The phasor voltage at each port can be calculated
as before, $\hat V_{i}=\int \dif x \dif y u_i\hat
V_T\equiv<u_{i}\hat{V}_{T}>$ and is linearly related to the phasor
currents $\hat{I}_j$ through the impedance matrix,
$\hat{V}_i=\sum_j Z_{ij} \hat{I}_j$.

To obtain an expression for the matrix $Z$, we expand $\hat{V}_T$
as before \cite{paperpart1} in the basis $\phi_n$, the eigenfunctions of
the closed
cavity. The result is
\begin{equation}
Z=-\mj kh\eta_0\sum_n\frac{\Phi_n \Phi_n^T}{k^2-k_n^2},
\label{eq:zcav3}
\end{equation}
where the vector $\Phi_n$ is [$\langle u_1 \phi_n \rangle$,
$\langle u_2 \phi_n \rangle$,...,$\langle u_M \phi_n \rangle]^T$.
 Using the random eigenfunction hypothesis, we write $\phi_n$ as a
 superposition of random plane waves. Thus the elements of the
 $M$-dimensional vector $\Phi_n$ will be Gaussian random
 variables with zero mean. Elements of $\Phi_n$ with different values of
$n$ corresponding to different eigenfunctions will be independent.
 However, for a given eigenfunction the elements of $\Phi_n$ may
 be correlated. This will be true, particularly, if two ports are
 close together, because the random superposition of plane waves leads
to an autocorrelation function $J_0(k\delta r)$ at two positions
separated  by $\delta r$ \cite{alhassid95}. To treat correlations we write
\begin{equation}
\Phi_n=L(k_n)w_n, \label{eq:utow}
\end{equation}
where $L$ is a non-random, as yet unspecified, $M\times M$ matrix
that depends on the specific coupling geometry at the ports and
may depend smoothly on $k_n$, and $w_n$ is an $M$-dimensional
Gaussian random vector with zero mean and covariance matrix equal to the
$M\times M$ identity matrix.
That is we require that the components of the random vector $w_n$ are
statistically
independent of each other, each with unit variance. Correlations between
ports are described by the off-diagonal components of $L$. The idea
behind (\ref{eq:utow}) is that the excitation of the ports by an
eigenmode will depend on the port geometry and on the structure of
the eigenmode in the vicinity of the ports.  The dependence on the
specific port geometry is not sensitive to small changes in the
frequency or cavity configuration and is embodied in the matrix
quantity $L(k)$.  The structure of the eigenmode in the vicinity
of the ports, however, is very sensitive to the frequency and
cavity configuration, and this motivates the statistical treatment
via the random plane wave hypothesis.  From the random plane wave
hypothesis, the excitation of the port from a given plane wave
component is random, and, since many such waves are superposed, by
the central limit theorem, the port excitation is a Gaussian
random variable, as reflected by the vector $w_n$.  In Paper I, we
have derived a result equivalent to (\ref{eq:utow}) for the case
of a one-port with a specific model of the excitation at the port
(namely, a vertical source current density $Iu(x,y)\hat z$ between
the plates). Our derivation here will be more general in that it
does not depend on a specific excitation or on the two-dimensional
cavity configuration used in Paper I.  Thus this derivation
applies, for example, to three dimensional cavities, and arbitrary
port geometries. From (\ref{eq:zcav3}) and (\ref{eq:utow}) we have
for the $Z$ matrix
\begin{equation}
Z=-\mj kh\eta_0\sum_n\frac{L(k_n)w_n w_n^TL^T(k_n)}{k^2-k_n^2}.
\label{eq:zcav4}
\end{equation}

We now take the continuum limit of (\ref{eq:zcav4}) and average over
$w_n$,
\begin{equation}
\langle Z \rangle =-\mj \int_0^{\infty} kh\eta_0 L(k')
\frac{\langle w_n w_n^T\rangle}{k^2-(k')^2}L^T(k')\frac{dk'^2}{\Delta },
\label{eq:zaverage}
\end{equation}
where $\Delta$ is the averaged spacing in $k_n^2$ values. We note
that the continuum limit is approached as the size of the cavity
is made larger and larger, thus making the resonance spacing
$(k^2_{n+1}-k^2_n)$ approach zero. Thus, the continuum limit
corresponds to moving the lateral walls of the cavity to infinity.
Using our previous one-port argument as a guide, we anticipate
that, if the pole in Eq.~(\ref{eq:zaverage}) at $k'^2=k^2$ is
interpreted in the causal sense (corresponding to outgoing waves
in the case with the walls removed to infinity), then $\langle Z
\rangle$ in (\ref{eq:zaverage}) is the radiation impedance matrix,
\begin{equation}
\langle Z \rangle =Z_R(k)=\hat{R}_R(k)+\mj \hat{X}_R(k),
\label{eq:corresp3}
\end{equation}
where $\hat V=Z_R(k)\hat I$ with $\hat V$ the $M$-dimensional
vector of port voltages corresponding to the $M$-dimensional
vector of port currents $\hat I$, in the case where the lateral
walls have been removed to infinity. With the above interpretation
of the pole, the real part of Eq.~(\ref{eq:zaverage}) yields
\begin{equation}
\hat{R}_R(k)=\pi kh\eta_0 L(k) L^T(k)/\Delta . \label{eq:RRk}
\end{equation}
Thus, Eq.~(\ref{eq:zcav3}) becomes
\begin{equation}
Z=-\frac{\mj}{\pi}\sum_n \Delta \frac{\hat{R}_R^{1/2}(k_n) w_n w_n^T
\hat{R}_R^{1/2}(k_n)}{k^2-k_n^2}, \label{eq:zfinal}
\end{equation}
where $\langle w_n w_n^T\rangle=1_M$.  (Note that the formula for $\Delta
$ is
different in two and three dimensions.)   In the case of
transmission line inputs that are far apart, e.g., of the order of
the cavity size, then the off-diagonal elements of $Z_R$ are small
and can be neglected. On the other hand, this will not be the case
if some of the transmission line inputs are separated from each
other by a short distance of the order of a wavelength. Similarly,
if there is a waveguide input to the cavity where the waveguide
has multiple propagating modes, then there will be components of
$\hat V$ and $\hat I$ for each of these modes, and the
corresponding off-diagonal elements of $Z_R$ for coupling between
these modes will not be small.

For the remainder of the paper, we will assume identical
transmission line inputs that are far enough apart that we may
neglect the off-diagonal elements of $Z_R$. As before, we will
take the eigenvalues $k_n^2$ to have a distribution generated by
RMT. Because the elements of $Z$ depend on the eigenvalues
$k_n^2$, there  will be correlations among the elements. In the
lossless case the
elements of the $Z$ matrix are imaginary, $Z=jX$, where $X$ is a
real symmetric matrix. Consequently $X$ has real eigenvalues. We
will show in Sec.~III that the distribution for individual
eigenvalues of $X$ is Lorentzian with mean and width determined by
the corresponding radiation impedance.

\subsection{Effects of Time-Reversal Symmetry
Breaking (TRSB)}

In the time-reversal symmetric system, the eigenfunctions of the
cavity are real and correspond to superpositions of plane waves
with equal amplitude waves propagating in opposite directions as
in Eq.~(13) of paper \cite{paperpart1}, which is recalled as
follows
\begin{equation}
\phi_n=\lim_{N\rightarrow
\infty}\sqrt{\frac{2}{AN}}Re\{\sum_{i=1}^{N}\alpha_i \exp(\mj k_n
\vec{e}_i\cdot \vec{x}+\mj\theta_i )\}, \label{eq:superposi}
\end{equation}
where $\alpha_i$, $\theta_i$ and $\vec{e}_i$ are random variables.
If a non-reciprocal element (such as a magnetized ferrite) is
added to the cavity, then time reversal symmetry is broken (TRSB).
As a consequence, the eigenfunctions become complex.
Eq.~(\ref{eq:superposi}) is modified by removal of the operation
of taking the real part, and the $\langle u \phi_n\rangle$ in
Eq.~(12) of paper \cite{paperpart1} also become complex. 
In practice, there exists a crossover
regime for the transition from situations where time reversal symmetry
applies to those it is fully-broken. An interested reader might refer
to situations where 
discussion in Ref.~\cite{anlage00} and the references therein.
However, in this paper, we will discuss only the case when
time-reversal symmetry is fully broken.
In this case we
find
\begin{equation}
\langle u_\ell \phi_n \rangle=[\Delta \hat{R}_R(k_n)]^{1/2}w_{\ell n}
\end{equation}
where $w_{\ell n}=(w_{\ell n}^{(r)}+\mj w_{\ell
n}^{(i)})/\sqrt{2}$ and $w_{\ell n}^{(r)}$ and $w_{\ell n}^{(i)}$
are real, independent Gaussian random variables with zero mean and
unit variance. The extra factor of $\sqrt{2}$ accounts for the
change in the normalization factor in Eq.~(\ref{eq:superposi}),
required when the eigenfunctions become complex. Further,
transpose $w^T_n$, in Eq.~(\ref{eq:zfinal}) is now replaced by the
conjugate transpose $w_n^\dag $.

A further consequence of TRSB is that the distribution of
eigenvalues is changed. The main difference is the behavior of
$P(s)$ for small $s$. In particular, the probability of small
spacings in a TRSB system ($P(s)\sim s^2$) is less than than of a
TRS system  ($P(s)\sim s$). 

For the sake of simplicity, we will assume all the transmission
lines feeding the cavity ports  are identical, and have the same
radiation  impedance, $Z_R=\hat{R}_R+\mj \hat{X}_R=(R_{R}+\mj
X_{R})1_M$, where $R_R$ and $X_R$ are real scalars. Analogous to the one
port case, we can define a model normalized reactance matrix
$\xi_{ij}=X_{ij}/R_R$ for the case $R_R(k_n)$ constant for $n\le
N$ and $R_R(k_n)=0$ for $n>N$,
\begin{equation}
\xi_{ij}=-\frac{1}{\pi}\sum_{n=1}^{N}\frac{w_{in} w^*_{jn}}{\tilde{k^2}
-\tilde{k_n^2}}, \label{eq:xgue}
\end{equation}
where $\tilde k^2=k^2/\Delta $, $w_{\ell n}=(w_{\ell n}^{(r)}+j
w_{\ell n}^{(i)})/\sqrt{2}$, $w_{\ell n}^{(r)}$ and
 $w_{\ell n}^{(i)}$ are real, independent Gaussian random variables with zero
mean and unit variance, $E(w_{in}^{*}w_{jn})=\delta_{ij}$.  Note
that a unitary transformation, $\xi '=U\xi U^\dag $, returns
(\ref{eq:xgue}) with $w_{in}$ and $w_{jn}$ replaced by $w'_{in}$
and $w'_{jn}$ where $w'_n=Uw_n$.  Since a unitary transformation
does not change the covariance matrix,
$E(w_{in}w^*_{jn})=E(w'_{in}w'^*_{jn})=\delta _{ij}$, the
statistics of $\xi $ and of $\xi '$ are the same; i.e., their
statistical properties  are invariant to unitary transformations.

\section{Properties of the Impedance Matrix and Eigenvalue Correlations
For Lossless Cavities}
The  universal fluctuation properties of the $Z$ matrix can be described
by
the model matrix $\xi_{ij}$ specified in  Eq.~(\ref{eq:xgue}). In
the TRS case the $w_{jn}$ are real Gaussian random variables with
zero mean and unit width and the  spacings satisfy Eq.~(16) in
Ref.~\cite{paperpart1}.  In the TRSB case the
$w_{jn}$ are complex and the spacings
between adjacent $k_n^2$ satisfy Eq.~(17) in Ref.~\cite{paperpart1}.

In the case under consideration of multiple identical ports,
$\xi_{ij}$ will have a diagonal mean part $\bar{\xi}\delta_{ij}$
for which all the diagonal values are equal. The eigenfunctions of
$\xi_{ij}=\bar{\xi}\delta_{ij}+\tilde{\xi}_{ij}$ and of its
fluctuating part $\tilde{\xi}_{ij}$ will thus be the same.
Consequently, we focus on the eigenvalues of the fluctuating part.

\begin{figure}
\includegraphics[scale=1]{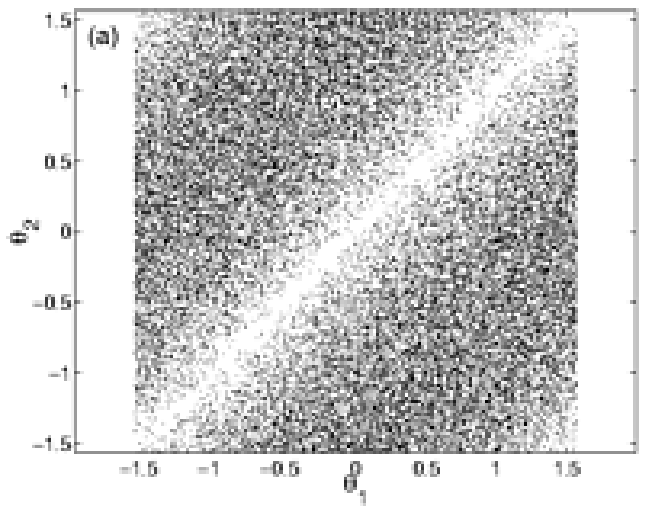}
\includegraphics[scale=1]{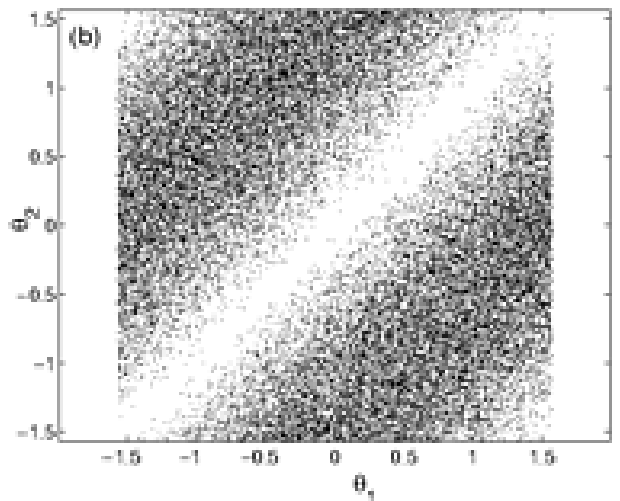}
\includegraphics[scale=1]{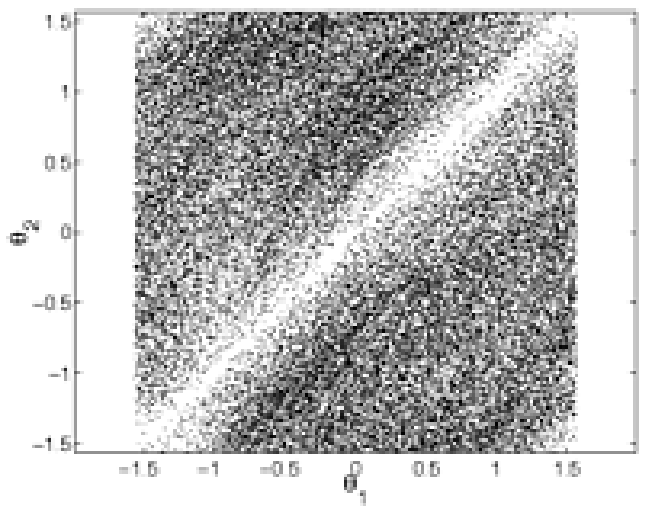}
\caption{(a) Scatter plot of $\theta_1$ vs $\theta_2$, in the TRS
case. (b) Scatter plot of $\theta_1$ vs $\theta_2$ in the TRSB
case.(c) Scatter plot of $\theta_1$ vs $\theta_2$ from the HFSS
simulation, with 100 realizations and sweeping frequency from
6.75GHz to 8.75GHz.} \label{fig:joint}
\end{figure}

We initially restrict our considerations to  the two-port case. We
recall that for the lossless one-port case there is no difference
in the statistics of the normalized impedance $\xi$ for the TRS and TRSB
cases. In both cases , it is Lorentzian with unit width. In the
lossless two-port case, however, essential differences are
observed when time reversal is broken.  Using (\ref{eq:xgue}), we
generate $10^6$ realizations of the 2 by 2 matrix $\xi $ in both
the TRS and TRSB cases, again for $N=2000$ and $k^2=1000$. In this
test we generated spectra based on an independent spacings
\cite{paperpart1}. For each realization we compute the eigenvalues
of the $\xi$ matrix. We find that in both the TRS and TRSB cases the
eigenvalues of the $\xi$-matrix are Lorentzian distributed  with unit
width. That is, histograms of the eigenvalues generated according to the 
TRS and TRSB prescriptions are identical. 
However, if we
consider the joint probability density function (PDF) of the two
eigenvalues for each realization, then differences between the TRS and
TRSB cases
emerge. We map the two eigenvalues $\xi_i$, $i=1$ or $2$, into the
range $[\pi/2,\pi/2]$ via the transformation $\theta_i={\rm
arctan}(\xi_i)$. Scatter plots of $\theta_2$ and $\theta_1$ for
$10^6$ random numerical realizations of the $\xi$ matrix are shown
in Fig.~\ref{fig:joint}(a) for the TRS case and in
Fig.~\ref{fig:joint}(b) for the TRSB case. The white diagonal band
in both cases shows that the eigenvalues avoid each other (i.e.,
they are anti-correlated). This avoidance is greater in the TRSB
case  than in the TRS case. The correlation,
\begin{equation}corr(\theta_1,\theta_2)\equiv\frac{\langle \theta_1
\theta_2 \rangle-\langle \theta_1 \rangle \langle \theta_2
\rangle}{\sqrt{\langle \theta_1^2 \rangle \langle \theta_2^2 \rangle}},
\label{eq:corre}
\end{equation}
is numerically determined to be -0.216 for the TRS case and -0.304 for
the TRSB case.

>From the construction of the $\xi$ matrices for the TRS and TRSB
cases their statistical properties are invariant under orthogonal
and unitary transformations, respectively. Random matrix theory
has been used to study these rotation-invariant ensembles and
predicts the joint density function of $\theta_1$ and $\theta_2$
\cite{theo4} to be,
\begin{equation}
P_{\beta}(\theta_1,\theta_2)\propto|\me
^{\mj 2\theta_1}-\me ^{\mj2\theta_2}|^{\beta},
\label{eq:jointpdf}
\end{equation}
where $\beta=1$ for the TRS case and $\beta=2$ for the TRSB case.
Note that based on Eq.~(\ref{eq:jointpdf}), the probability
density function for one of the angles $P(\theta_1)=\int d\theta_2
P(\theta_1,\theta_2)$ is uniform. From the definition
$\theta=\arctan \xi$, this is equivalent to the eigenvalues of the
$\xi$ matrix having Lorentzian distributions $(P_\xi (\xi
_i)=P_\theta (\theta _i)|d\theta _i/d\xi _i|=|d\theta _i/d\xi
_i|/2\pi )$.

The correlation coefficients calculated from the numerical results
in Figs.~\ref{fig:joint}(a) and \ref{fig:joint}(b)  are consistent with
the predictions of the
random matrix theory from Eq.~(\ref{eq:jointpdf}), that is, -0.216
for the TRS case and -0.304 for the TRSB case. This implies that the
distribution of spacings and the long range correlations in the
eigenvalues of the random matrix, which are ignored in the
construction of the $k_n^2$ in the above computation
are not
important in describing the statistics of {\it  lossless} impedance
matrices. As we have discussed in \cite{paperpart1},
these correlations could be included using a sequence of $k_n^2$
generated
by the eigenvalues of a random matrix. (We note that \cite{paperpart1},
lossy cavities yield statistics that are different in the TRS and TRSB
cases.)

Now we test these predictions for numerical simulations of
the chaotic cavity considered in paper \cite{paperpart1}. We use the
HFSS software to calculate the cavity impedance matrix and
radiation impedance matrix for a 2-port case. We locate the two
ports, at the positions ($x$, $y$)=(14cm, 7cm) and ($x$,
$y$)=(27cm, 13.5cm). As in \cite{paperpart1}, we also include the 0.6 cm
cylindrical
perturbation which is located alternately at 100 random points in
the cavity, and we numerically calculate the impedance matrix  for
4000 frequencies in the range 6.75GHz to 8.75GHz. We obtain a
normalized $Z$ matrix, which is analogous to the $\xi$ matrix
defined in Eq.~(\ref{eq:xgue}) according to
\begin{equation}
\xi_{hfss}=R_R^{-1}(Im[Z_{cav}]-{1}_2 X_R),
\label{eq:znorm_2port}
\end{equation}
where $1_2$ is the 2 by 2 identity matrix, $Z_{cav}$ is the  2 by 2
impedance matrix calculated by
HFSS, and $X_R$ and $R_R$ are the radiation reactance and
resistance for a single port. For each realization of
$\xi_{hfss}$ we calculate its eigenvalues $\xi_i=\tan
\theta_i$,
$i=1,2$, and plot the values on the $\theta_1$ vs. $\theta_2$
plane, as shown in Fig.~\ref{fig:joint}(c). The anti-correlation of the
angles is seen in the figure, and $corr(\theta_1, \theta_2)$ from
(\ref{eq:corre}) is
-0.205, which is comparable with what we expect for the TRS case, -0.216.

\begin{figure}
\includegraphics[height=1in,width=1.3in] {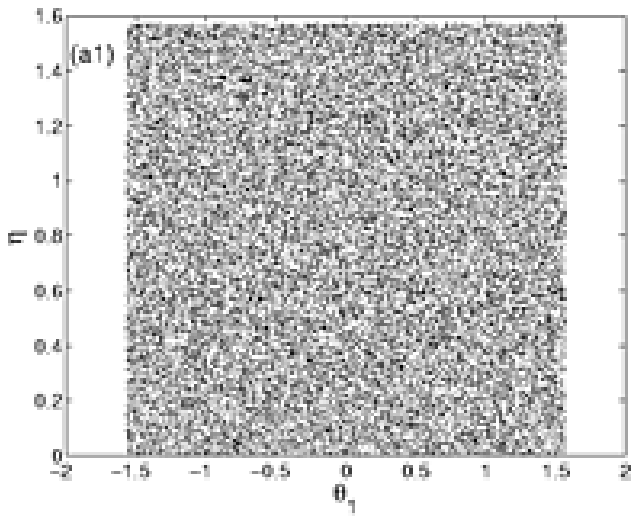}
\includegraphics[height=1in,width=1in]{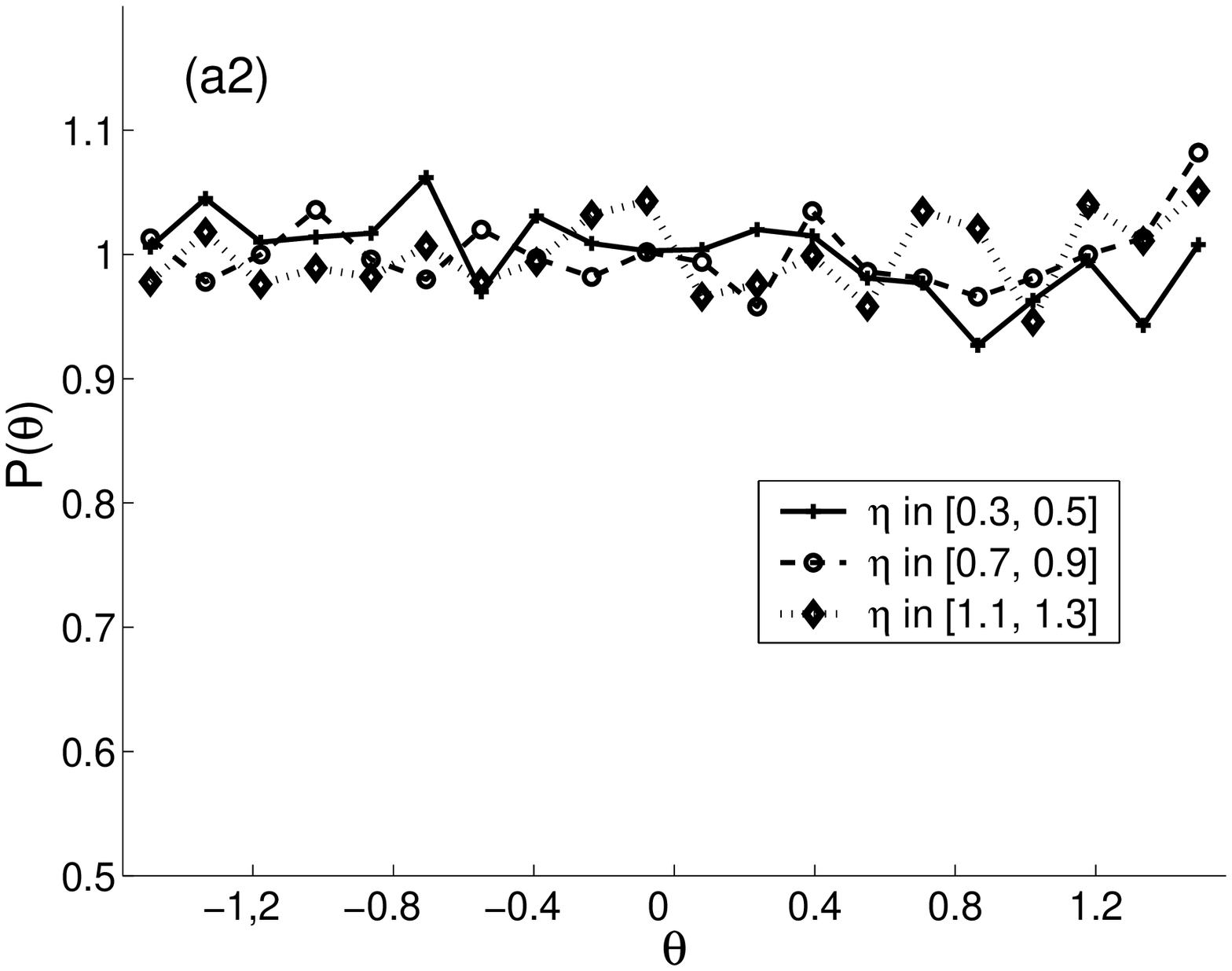}
\includegraphics[height=1in,width=1in]{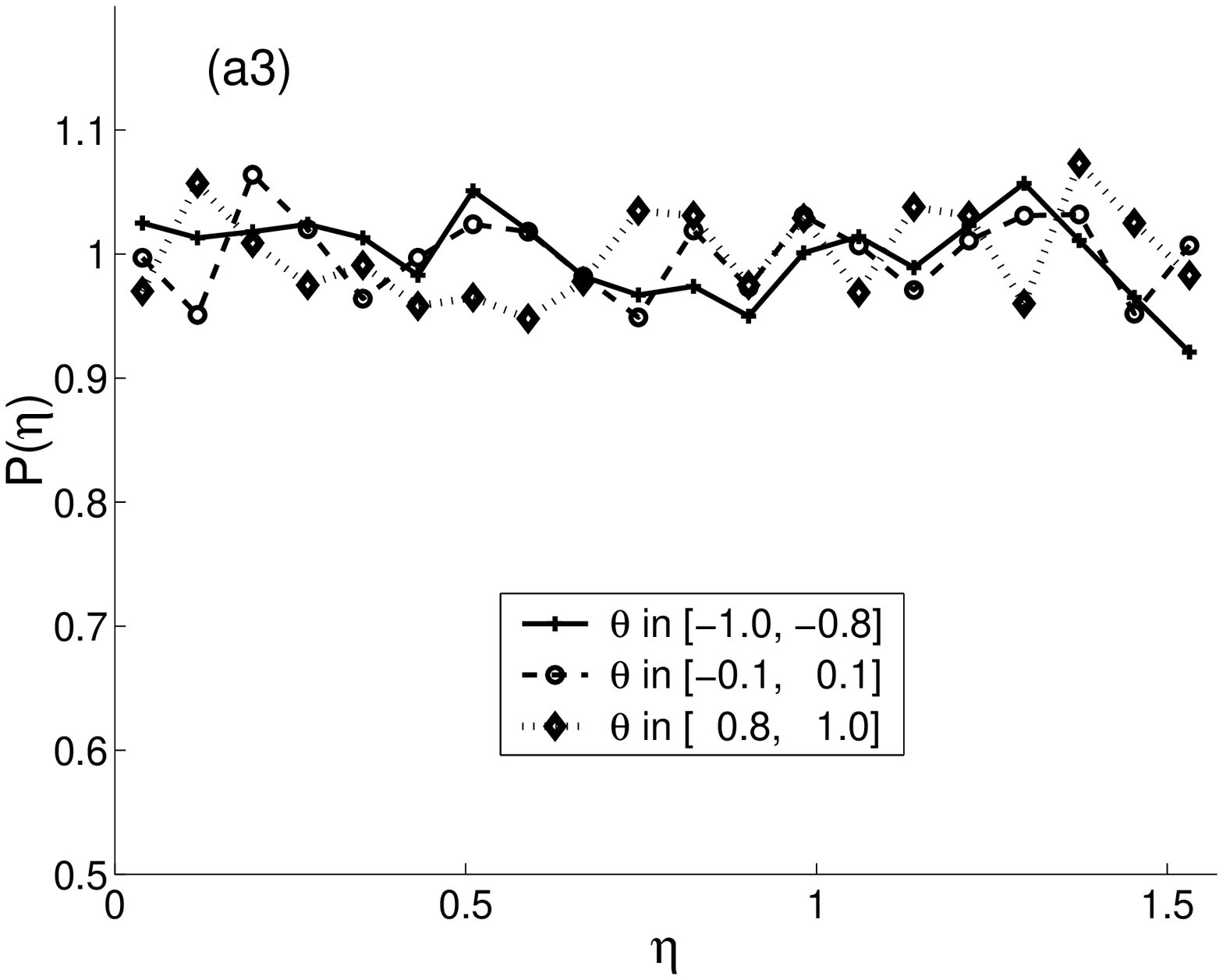}
\includegraphics[height=1in,width=1.3in]{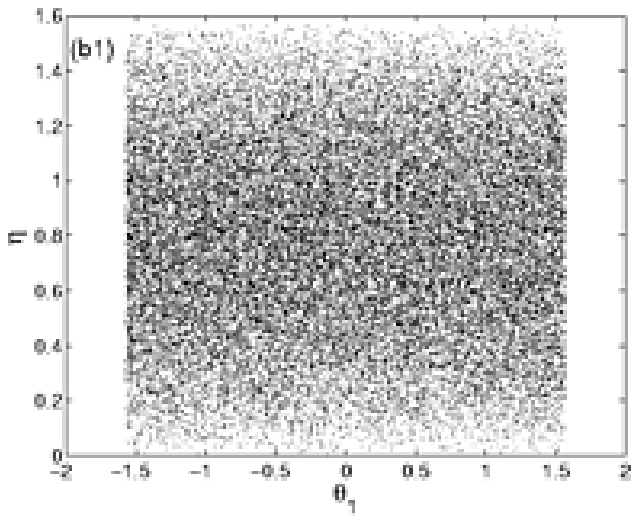}
\includegraphics[height=1in,width=1in]{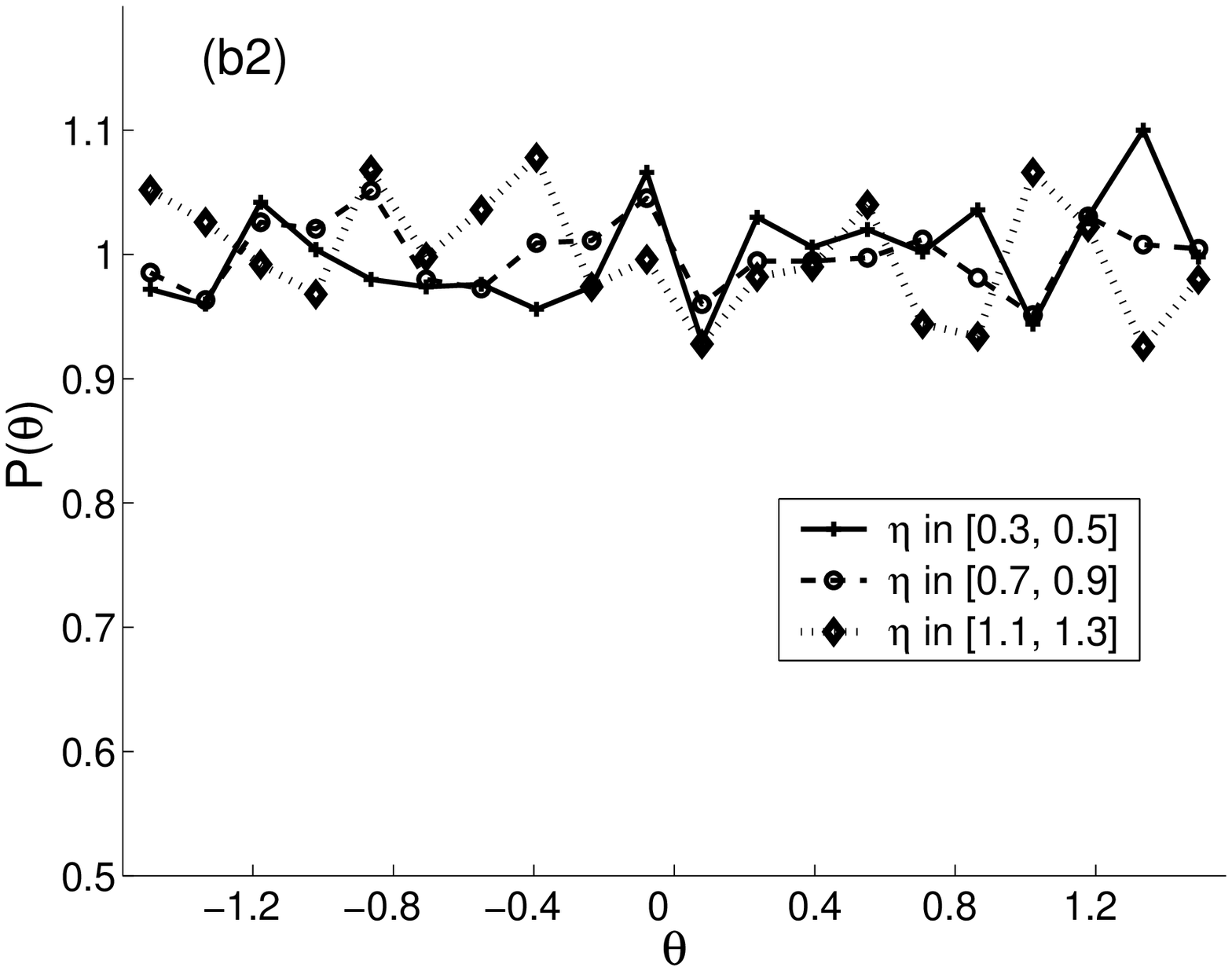}
\includegraphics[height=1in,width=1in]{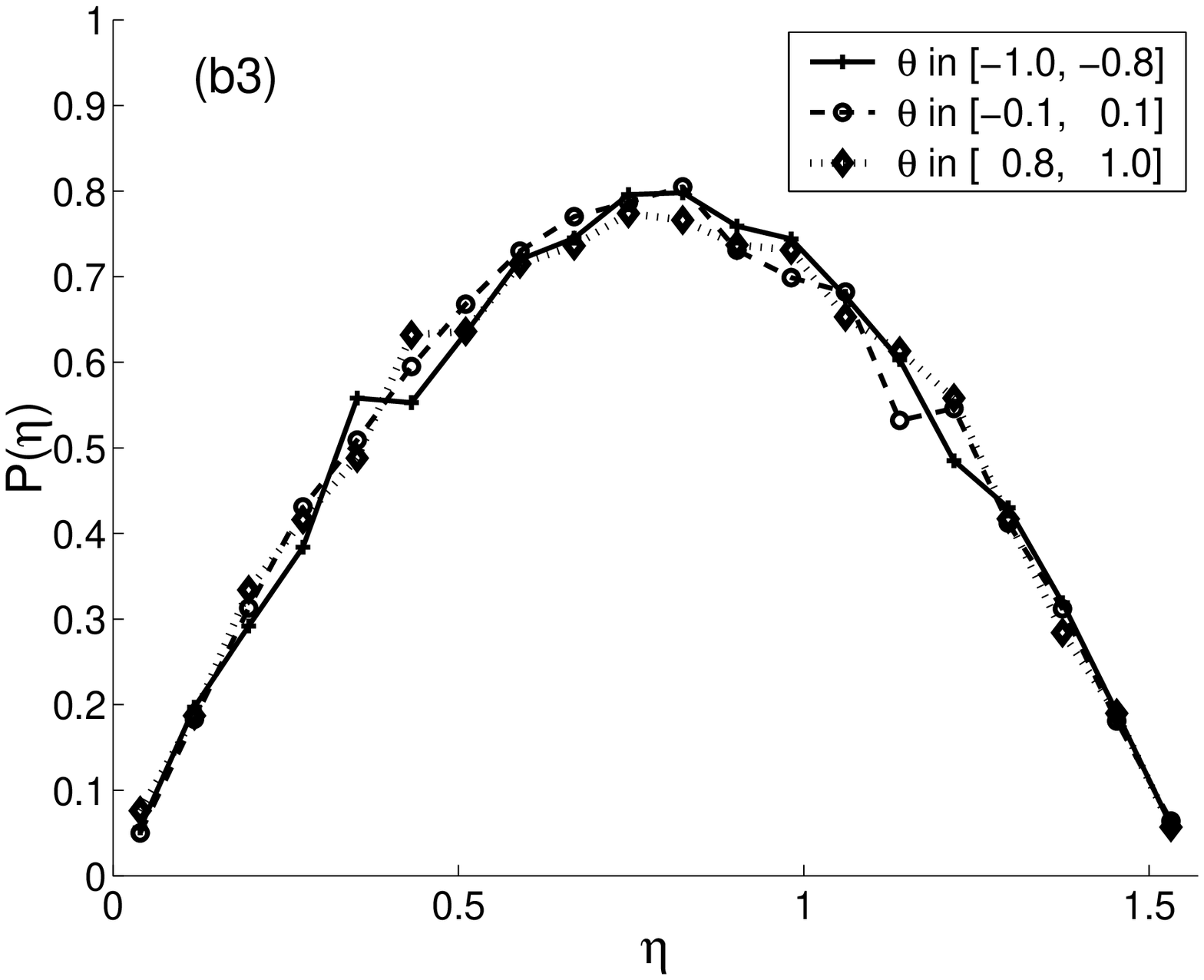}
\includegraphics[height=1in,width=1.3in]{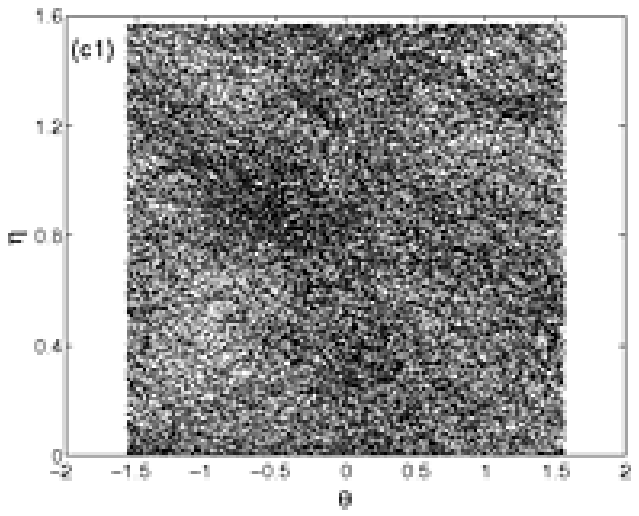}
\includegraphics[height=1in,width=1in]{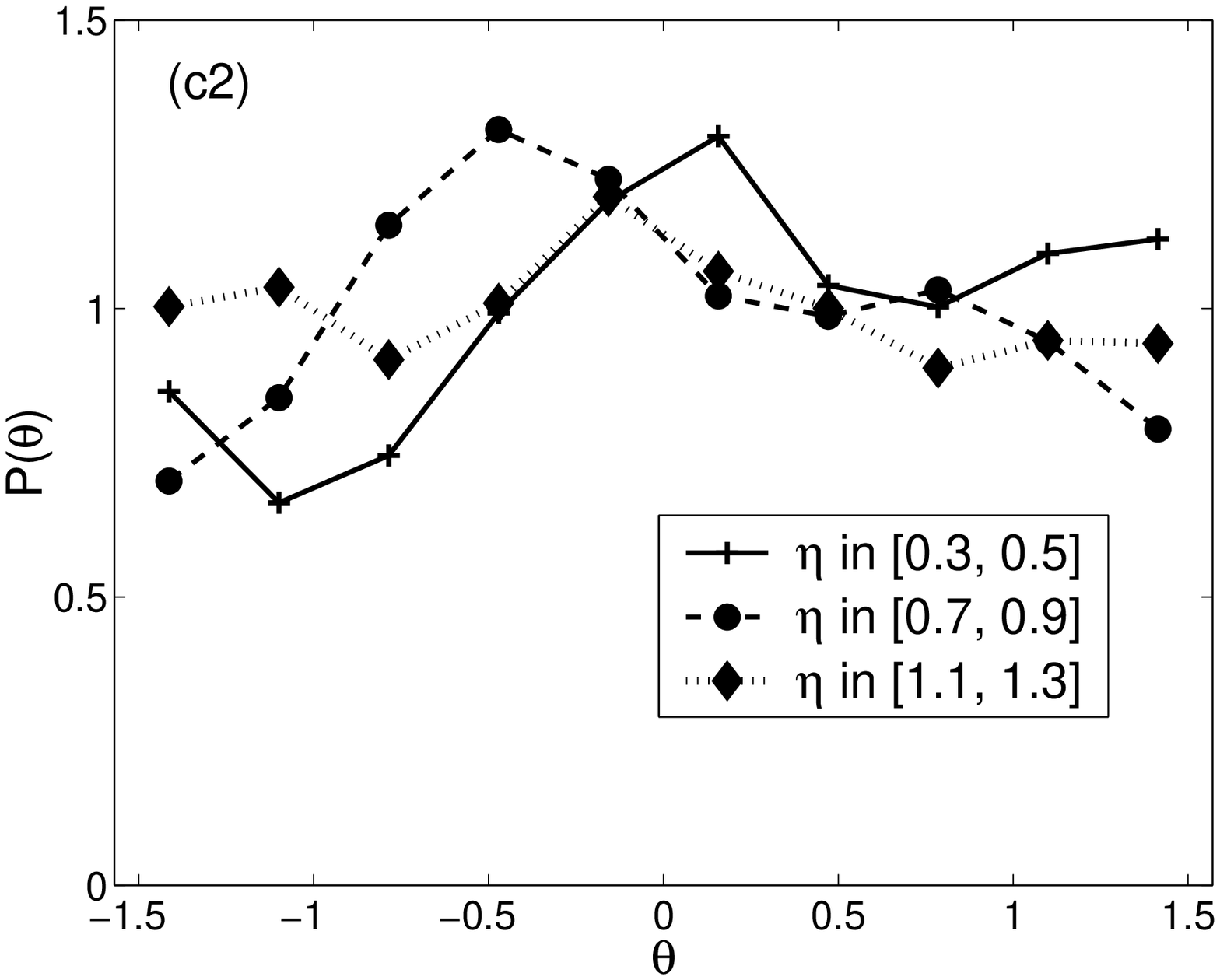}
\includegraphics[height=1in,width=1in]{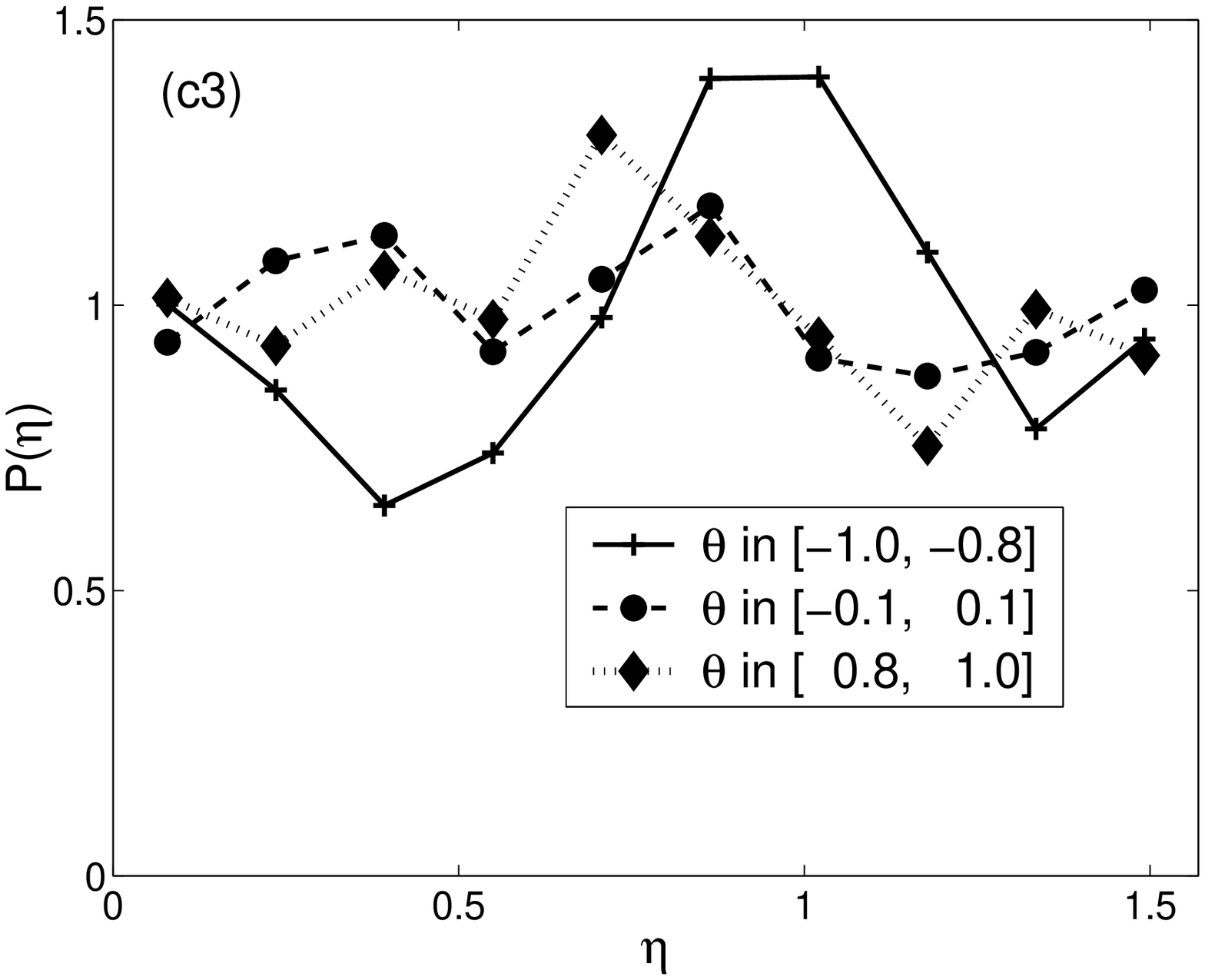}
\caption{Scatter plot of $\eta$ vs $\theta$ for (a1) the model impedance
in the TRS case,
(b1) the model impedance in the TRSB case, and (c1) from the HFSS
simulation.  Plots (a2)
and (a3) [(b2) and (b3), (c2) and (c3)] show conditional probability
for
$\theta $ and for $\eta $ for the model TRS case [model TRSB case, the
HFSS simulation].}
\label{fig:thetayita}
\end{figure}

So far we have focused on the eigenvalues of the impedance
matrix. The eigenvectors of $Z$ are best described in terms of the
orthogonal matrix whose columns are the orthonormal eigenfunctions
of $Z$. Specially, in the TRS case, since $\xi$ is real and symmetric,
\begin{equation}
  \xi=O
  \begin{pmatrix}
    \tan \theta_1 & 0 \\
    0 & \tan \theta_2 \
  \end{pmatrix}O^T,
\label{eq:xieigen_trs}
\end{equation}
where $O^T$ is the transpose of $O$,
and $O$ is an orthogonal matrix, which we express in the form
\begin{equation}
  O=
  \begin{pmatrix}
    \cos\eta & \sin \eta \\
    -\sin \eta & \cos \eta \
  \end{pmatrix}.
\label{eq:Opara}
\end{equation}
A scatter plot representing the joint pdf of the angle $\eta$ and one
of the eigenvalue angles
$\theta_1$ is shown in
Fig.~\ref{fig:thetayita}(a1). In analogy to how we obtain the realizations
used in Fig.~2 in \cite{paperpart1}, this plot is obtained by
inserting random
choices for the $k_n^2$ and $w_{in}$ in (\ref{eq:xgue}).
Notice that we have restricted $\eta$ in Fig.~\ref{fig:thetayita}(a1)
to the range $0\leq \eta \leq \pi /2$.  This can be justified as
follows. The columns of the matrix $O$ in (\ref{eq:Opara}) are the
eigenvectors of $\xi $.  We can always define an eigenvector such
that the diagonal components of $O$ are real and positive.
Further, since the eigenvectors are orthogonal, one of them will
have a negative ratio for its two components.  We pick this one to
be the first column and hence this defines which of the two
eigenvalues is $\theta _1$.  The scatter plots in
Fig.~\ref{fig:joint} show that the restriction on $\eta $
maintains the symmetry of $\theta _1$ and $\theta _2$, vis.
$P_\beta (\theta _1,\theta _2)=P_\beta (\theta _2,\theta _1)$.
Also in the Fig.~\ref{fig:thetayita}(a2) (and (a3)), we plot the
conditional distribution of $\theta$ (and $\eta$) for different
values of $\eta$ (and $\theta$). As can be seen, these plots are
consistent with $\eta$ and $\theta$ being independent. This is
also a feature of the random matrix model \cite{mehta91}. This
independence will be exploited later when the $S$ matrix is
considered.

For TRSB systems, the $\xi$ matrix is Hermitian $\xi^T=\xi^*$. A
unitary matrix of eigenvectors that diagonalizes it can be
parameterized as
\begin{equation}
  U=
  \begin{pmatrix}
    \cos \eta & \sin\eta e^{\mi\zeta} \\
   -\sin\eta \me^{-\mi\zeta}  & \cos \eta \
  \end{pmatrix}.
\label{eq:Upara}
\end{equation}
Thus, there is an extra parameter $\zeta $ characterizing the
complex eigenvectors of the $\xi $ matrix. According to random
matrix theory, the eigenfunctions and eigenvalues are
independently distributed, i.e. $\eta$ in the $U$ matrix should be
independent of $\theta_1$, $\theta_2$. This expectation is
confirmed in Fig.~\ref{fig:thetayita}(b) where a scatter plot of
$\theta_1$ vs $\eta$ and conditional distributions of $\theta $
and of $\eta $ are shown.

Again, we test the independence of $\theta$ and $\eta$ with HFSS
calculations. Using the $\xi_{hfss}$ matrix obtained from
Eq.~(\ref{eq:znorm_2port}), the angles $\theta$ and $\eta$ can be
recovered from the eigenvalues and the eigenvectors of the
$\xi_{hfss}$. With the ensemble generated by sweeping the
frequency from 6.75-8.75GHz and considering 100 different
locations of our cylindrical perturber, we obtain the joint
distribution of $\theta$ and $\eta$ in
Fig.~\ref{fig:thetayita}(c1) as well as their individual
distributions  in Fig.~\ref{fig:thetayita}(c2) and (c3). Here we
see that the distributions are qualitatively similar to those of
the model impedance matrix in the TRS case. However, there are
significant departures which need to be investigated. It is likely
that these are the result of the same strong multi-path
interference which gave rise to the reactance variations in the
one port case shown in Paper \cite{paperpart1}.

\section{Averaged Reflection Coefficient in Multi-port Case}
In this section, we use our knowledge of the statistical
properties of the $Z$ matrix to deduce properties of
the $S$ matrix, particularly for the ensemble average of the
reflection coefficient $\langle |S_{11}|^2 \rangle$. For a system
with two ports, in the lossless case considered here we note $\langle
|S_{12}|^2 \rangle=1-\langle |S_{11}|^2 \rangle$

According to the previous section, for the case of non-perfect
coupling, model of the cavity impedance matrix can be expressed as
$Z=\hat{R}_R^{1/2}\xi \hat{R}_R^{1/2}+j\hat{X}_R$, where $Z_R$ is the
$2\times
2$ radiation impedance and $\xi$ is a $2\times 2$ random matrix
generated according to Eq.~(\ref{eq:xgue}). If the incoming
frequency is restricted in a narrow range, the radiation impedance
$Z_R$ is essentially constant. In this paper we assume that
identical ports are connected to identical transmission line,
i.e., $Z_R$ and the transmission line characteristic impedance
$Z_0$ are diagonal matrices with equal diagonal elements. Thus, we
obtain the expression for the $S$ matrix, $S=(Z+Z_0)^{-1}(Z-Z_0)$,
\begin{equation}
S=[(\gamma_R\xi+j\gamma_X1_2)+1_2]^{-1}[(\gamma_R\xi+j\gamma_X1_2)-1_2],
\label{eq:s_general}
\end{equation}
where $\gamma_R=R_R/Z_0$, $\gamma_X=X_R/Z_0$ are scalars and $1_2$
is the $2\times 2$ identity matrix . These two parameters, as we
show later, fully specify the coupling effects on the wave
transport process. The special case of perfect coupling
corresponds to $\gamma_R=1$ and $\gamma_X=0$.

\subsection{Lossless Two-port Case}
We recall that for  TRS systems the reactance matrix $X$ is real
and symmetric, and can be diagonalized by an orthogonal matrix
$O$, Eq.~(\ref{eq:Opara}). If identical ports are connected to
identical transmission lines of characteristic impedance $Z_0$,
then the scattering matrix $S$ is also diagonalized by $O$, and we
can write
\begin{equation}
  S=O
  \begin{pmatrix}
    \me^{\mj\phi_1} & 0 \\
    0 & \me^{\mj\phi_2} \
  \end{pmatrix}O^T.
\label{eq:seigen_trs}
\end{equation}
The scattering phases $\phi_1$ and $\phi_2$ are then related to
the eigenvalue angles $\theta_i$ by formulas analogous to the
one-port case, $\tan (\pi/2-\phi_i/2)=\gamma_R \tan
\theta_i+\gamma_X$.

Substituting Eq.~(\ref{eq:Opara}) for $O$ in (\ref{eq:seigen_trs}) and
multiplying the matrices, we obtain
\begin{equation}
|S_{11}|^2=\cos^4\eta+\sin^4\eta+2\cos^2\eta\sin^2\eta\cos(\phi_1-\phi_2).
\label{eq:s11sqr_trs}
\end{equation}
We can now compute the expected value of the square of $|S_{11}|$
by assuming that $\eta$ is independent of the angles $\phi_1$ and
$\phi_2$ and is uniformly distributed, which yields $\langle
\cos^4\eta+\sin^4\eta \rangle =3/4$, $2\langle
\cos^2\eta\sin^2\eta \rangle =1/4$ and
\begin{equation}
\langle |S_{11}|^2 \rangle=\frac{3}{4}+\frac{1}{4}\langle
\cos(\phi_1 -\phi_2) \rangle .\label{eq:formulates11_goe}
\end{equation}
Assuming the angles $\theta_1$ and $\theta_2$ are distributed
according to Eq.~(\ref{eq:jointpdf}) and using the relation
between $\phi_{1,2}$ and $\theta_{1,2}$, evaluation of $\langle
\cos(\phi_1-\phi_2)\rangle$
 is carried out in Appendix. The result is
\begin{equation}
\langle |S_{11}|^2 \rangle
=1-\frac{1-|\rho_R|^4}{8|\rho_R|^2}-\frac{(1-|\rho_R|^2)^3}
{16|\rho_R|^3}\ln \frac{1-|\rho_R|}{1+|\rho_R|},
\label{eq:results11_goe}
\end{equation}
where the ``the free space reflection coefficient" $\rho_R$ is defined as
the same way in the Paper \cite{paperpart1},
\begin{equation}
\rho_R=|\rho_R|e^{j\phi_R}=\frac{\gamma_R+j\gamma_X-1}{\gamma_R+j\gamma_X+1}.
\label{eq:rhor}
\end{equation}

We first check the asymptotic behavior for the power transmission
coefficient $T=1-|S_{11}|^2$ implied by the formula
(\ref{eq:results11_goe}). In the non-coupled case, $|\rho_R|=1$, i.e.,
all the incoming power is reflected, and we obtain from
(\ref{eq:results11_goe}) $\langle T \rangle=0$. On the other hand,
in the perfect coupling case, $|\rho_R|=0$,
$\ln[(1+|\rho_R|)/(1-|\rho_R|)]$ in the (\ref{eq:results11_goe})
can be expanded as $2(|\rho_R|-|\rho_R|^3/3)$. Therefore, $\langle
T \rangle=1/3$. This is consistent with the result in
Ref.~\cite{ekogan}, $\langle R \rangle =2 \langle T \rangle$. That is,
in the perfect coupling case the average of the reflected power is
twice that of the transmitted.

\begin{figure}
\includegraphics[scale=0.3]{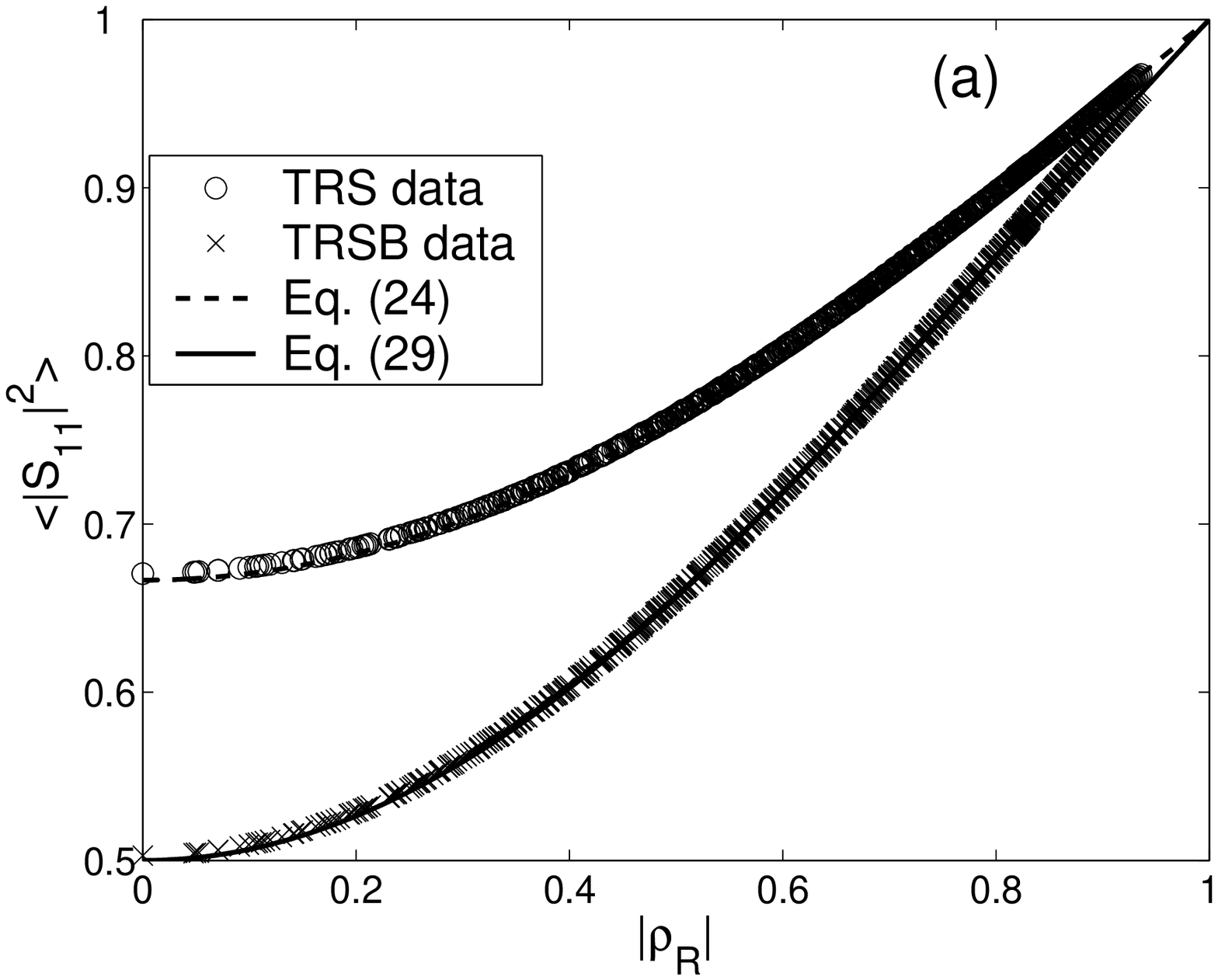}
\includegraphics[scale=0.3]{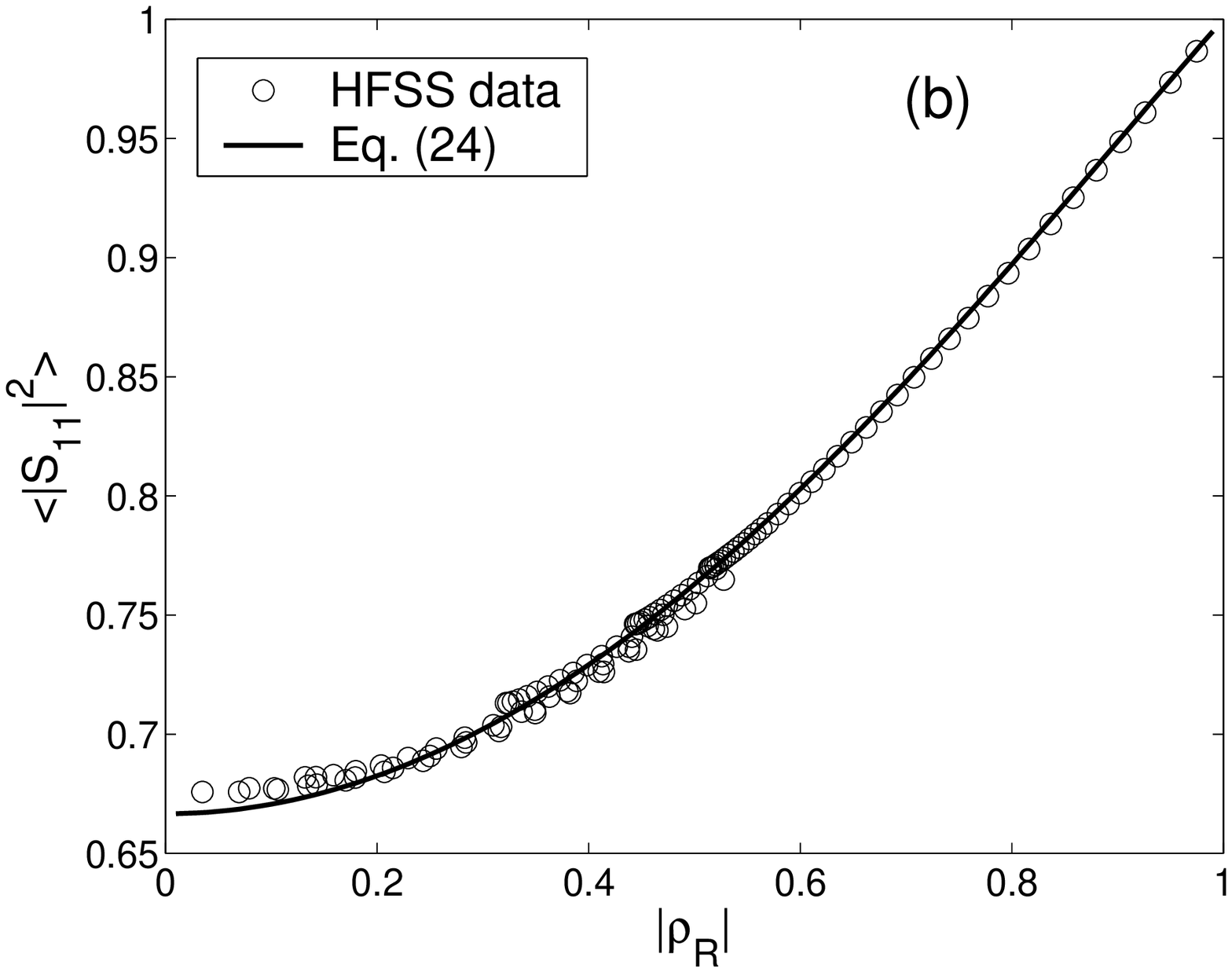}
\caption{(a) Numerical simulation for the average reflection
coefficient $\langle|S_{11}|^2\rangle $ vs magnitude of $\rho_R$
defined in
Eq.~(\ref{eq:rhor}) for the TRS and the TRSB system,
taking $10^6$ realization of the impedance matrix, 30 uniformly
spaced values of $\gamma_R$ from 0.1 to 3, and 31 equally spaced
values of $\gamma_X$ from 0 to 3. (b)  Average reflection
coefficient $\langle|S_{11}|^2\rangle $ vs $|\rho_R|$ using the cavity
impedance and radiation impedance from HFSS calculation and varying the
values of $Z_0$ and the capacitive reactance $Y$} \label{fig:collapse}
\end{figure}

Eq.~(\ref{eq:results11_goe}) shows that the averaged power
reflection and transmission coefficients only depend on the
magnitude of $\rho_R$ and not its phase.  A plot of $\langle
|S_{11}|^2 \rangle$ versus $|\rho_R|$ is shown in
Fig.~\ref{fig:collapse}(a). Also shown are data points obtained by
taking $10^6$ realizations of the impedance matrix (\ref{eq:xgue})
with eigenvalue statistics generated from TRS spectrum and
computing the average of $|S_{11}|^2$ for different combinations
of $\gamma_R$ and $\gamma_X$ characterizing the radiation
impedance. The data confirm that the average of $|S_{11}|^2$
depends only on the magnitude of the free space reflection
coefficient and not its phase.

In the TRSB case,  the eigenvalues of the $X$ matrix are still real, but the
eigenvectors are complex. In this case, Eq.~(\ref{eq:seigen_trs}) is
replaced by
\begin{equation}
 S=U
  \begin{pmatrix}
    \me^{\mj\phi_1} & 0 \\
    0 & \me^{\mj\phi_2} \
  \end{pmatrix}U^{\dag},
\label{eq:seigen_trsb}
\end{equation}
where the unitary matrix $U$ is given by Eq.~(\ref{eq:Upara}).
Multiplying the matrices in Eq.~(\ref{eq:seigen_trsb}), we find
the same expression  for $|S_{11}|^2$, Eq.~(\ref{eq:s11sqr_trs}),
as in the TRS case. The average of $|S_{11}|^2$ will be different
in the TRSB case because of the different statistics for $\eta$,
$\theta_1$ and $\theta_2$ which characterize the eigenfunctions
and eigenvalues of the impedance matrix. In particular, $\eta$ has
a distribution, arising from the SU(2) group \cite{su2group},
\begin{equation}
P_{\eta}(\eta)=|\sin(2\eta)|,
\end {equation}
which yields $\langle \cos^4\eta+\sin^4\eta \rangle =2/3$,
$2\langle \cos^2\eta\sin^2\eta \rangle =1/3$, thus,
\begin{equation}
\langle |S_{11}|^2 \rangle=\frac{2}{3}+\frac{1}{3} \langle
\cos(\phi_1-\phi_2) \rangle. \label{eq:formulates11_gue}
\end{equation}
Recalling that $\theta_1$ and $\theta_2$ are distributed according
to (\ref{eq:jointpdf}) with $\beta=2$, this results in a different
set of integrals (see Appendix). The result is
\begin{equation}
\langle |S_{11}|^2 \rangle=
1-\frac{(|\rho_R|^2-1)(|\rho_R|^2-3)}{6}, \label{eq:results11_gue}
\end{equation}
which  depends only on the magnitude of the free space reflection
coefficient. A plot of  $\langle
|S_{11}|^2 \rangle$ from Eq.~(\ref{eq:results11_gue})
versus
$|\rho_R|$ is also shown in Fig.~\ref{fig:collapse}(a), along with
data point obtained by taking $10^6$ realizations of the TRSB
impedance matrix (\ref{eq:xgue}) generating from random numbers and
computing
the average of
$|S_{11}|^2$ for different combinations of $\gamma_R$ and
$\gamma_X$ characterizing the free space impedance. Once again,
the data collapse to the curve predicted in
Eq.~(\ref{eq:results11_gue}).

We now test the relation between $\langle|S_{11}|^2\rangle $ and
$|\rho_R|$ with the impedance matrices we obtained  from the HFSS
two-port calculations. We can vary the transmission line impedance
$Z_0$ and generate $\langle|S_{11}|^2\rangle $ and $|\rho_R|$.
However, the range of $|\rho_R|$ values accessible doing this is
limited  because of the large inductive radiation reactance
associated with the coupling port. To extend the range of
$|\rho_R|$ we add a shunt susceptance $Y=(j\omega C)$ in
parallel
with each port. This results in a modified cavity impedance matrix
$Z_{cav}^{'}=(Z_{cav}^{-1}+j\omega C 1_2)^{-1}$. We then
form the scattering matrix
\begin{equation}
S=(Z_{cav}^{'}+Z_0)^{-1}(Z_{cav}^{'}-Z_0).
\label{eq:newS}
\end{equation}
The corresponding free space reflection coefficient is generated by
$Z_R^{'}=(Z_R^{-1}+j\omega C)^{-1}$ and
$|\rho_R|=|Z_R^{'}+Z_0|^{-1}|Z_R^{'}-Z_0|$.
By choosing appropriate
combinations of $\omega C$ and $Z_0$, we can achieve a range of
$|\rho_R|$ values between 0 and 1. For each $|\rho_R|$ value, we
average $|S_{11}|^2$ over frequencies and realizations and plot the
points on Fig.~\ref{fig:collapse}(b). These compare favorably with
the theoretical result (solid curve) based on the random matrix theory
results.

\subsection{M-port Case, $M > 2$}
Using the random coupling model (\ref{eq:xgue}) and assuming
perfect coupling $\gamma_R=1$, $\gamma_X=0$ (i.e. $|\rho_R|=0$),
we have simulated the $S$ matrix for cases of two to seven, 13 and 57
ports.
The results for the average reflection and transmission
coefficients were found to satisfy:
\begin{equation}
\text{TRS}: \qquad \langle|S_{ij}|^2\rangle=
  \begin{cases}
    \frac{2}{M+1} & \qquad i=j, \\
    \frac{1}{M+1} & \qquad i\neq j,
  \end{cases}
\end{equation}
and
\begin{equation}
   \text{TRSB}: \qquad
\langle|S_{ij}|^2\rangle=
  \begin{cases}
    \frac{1}{M} & \qquad i=j, \\
    \frac{1}{M} & \qquad i\neq j,
  \end{cases}
\end{equation}
where $M$ is the number of ports connecting the cavity to
transmission lines.  It seems that, in the TRS case, the input
waves ``remember" their entry port and have a preference for
reflection through it (this is related to the concept of ``weak
localization" reviewed in \cite{weakloca}). In contrast, for the
TRSB case, the waves behave as if they forget through which port
they entered the cavity, and thus all the ports have equal
probability of being the output for the waves.

It was shown by Brouwer and Beenakker \cite{brouwer97} that
scattering in multiport lossless systems can be related to that in
a single-port, lossy system. It was proposed that the
introduction of $N'$ ($N'>>1$) fictitious ports into the
scattering matrix of a lossless system would give equivalent statistics
for the reflection coefficient seen at a single port as would be
obtained for a single port model with a uniform internal loss.
Considering
a system with $M$ ports all perfectly matched, we can pick port 1
as the input and consider the other ports as a form of
dissipation. Due to the energy escaping from the other $(M-1)$
ports, we will obtain a reflection coefficient $S_{11}$ with
magnitude less than one, which is similar to that obtained in the
one-port lossy case (i.e., with losses due to finite wall
conductivity). The cavity impedance seen from port 1, $Z_1$, is
calculated from $S_{11}$, one of the elements from the $M$ by $M$
scattering matrix,
\begin{equation}
Z_1=R_R\frac{1+S_{11}}{1-S_{11}}+jX_R. \label{eq:z1}
\end{equation}
When normalized by the radiation impedance this corresponds to a
complex impedance $\zeta_M=(1+S_{11})/(1-S_{11})=\rho+j\xi$. On
the other hand, we can generate the lossy one-port impedance
$\zeta$ from Eq.~(48) in Ref.~\cite{paperpart1}, modelling the
lossy effect by adding a small imaginary term to the frequency
\cite{doron90}. We can then compare the statistics of $\zeta$ from
the lossy one port and $\zeta_M$ from the $M$-port lossless case (We note 
that approximate analytic formula for the distributions of the real and
imaginary parts of $\zeta$ have recently been given by Fyodorov and
Savin \cite{fyodorov04}). An appropriate value of the damping parameter in
the
one port case, $\tilde{k}^2\sigma$ ($\sigma=1/Q$), can be determined so
that the average value of $|S_{11}|^2$ in the lossy case is equal
to $2/(M+1)$ for the TRS case (or $1/M$ for the TRSB case). Then
we can compare the real and imaginary parts of the impedances
obtained in the  two different ways. In Fig.~4, we include the
results for the three different number of ports, $M$=4, 13 and 57,
and the corresponding one port result. For $M=4$ we note that the
distributions  are similar but clearly not the same. However, for
$M$=13 or 57, the distributions for $\zeta$ and $\zeta_M$ are much
closer. Thus, we confirm that distributed damping and a large
number of output channels are equivalent so as to affect the
distribution of the sub-unitary scattering matrix.
\begin{figure}
\includegraphics[scale=0.3]{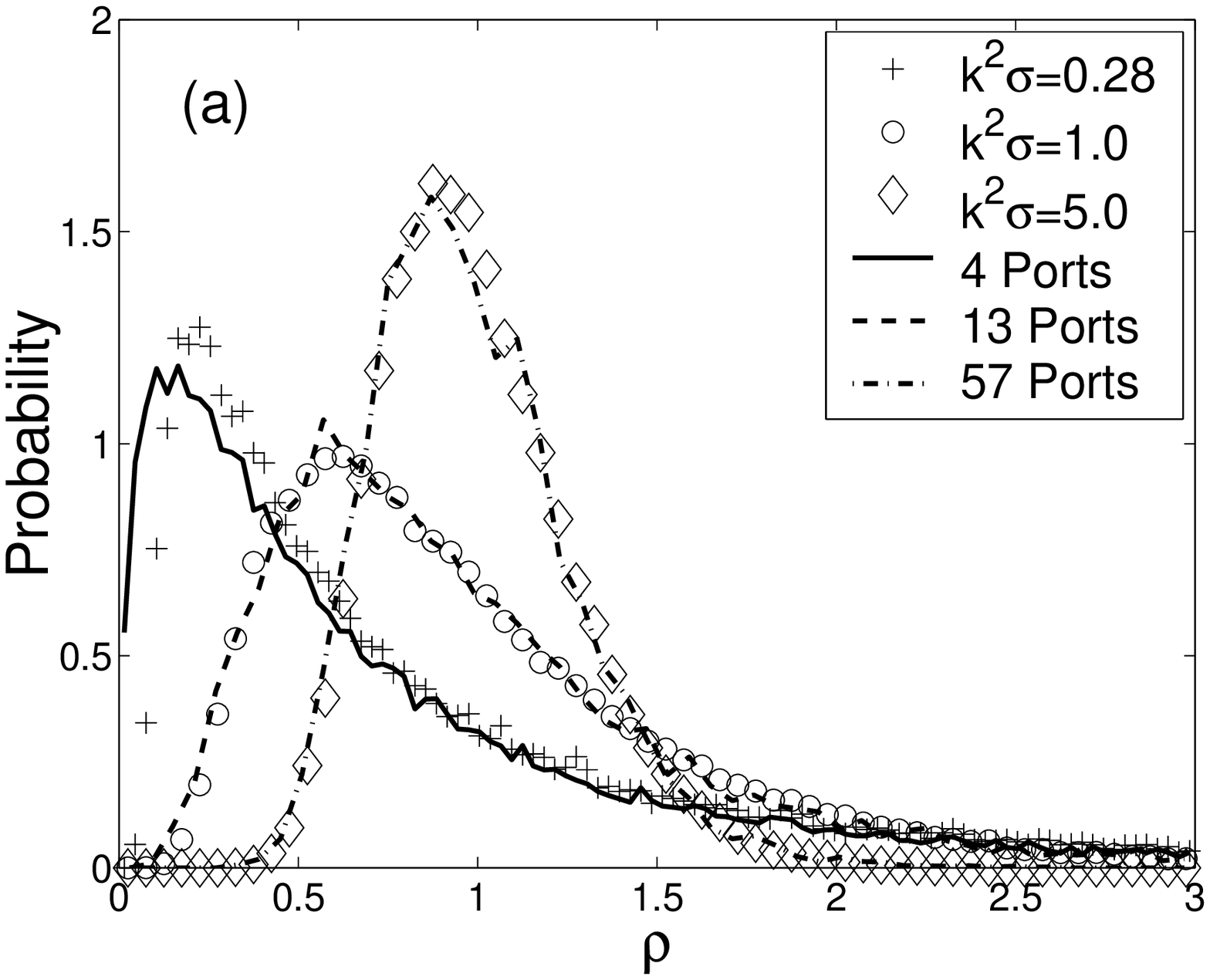}
\includegraphics[scale=0.3]{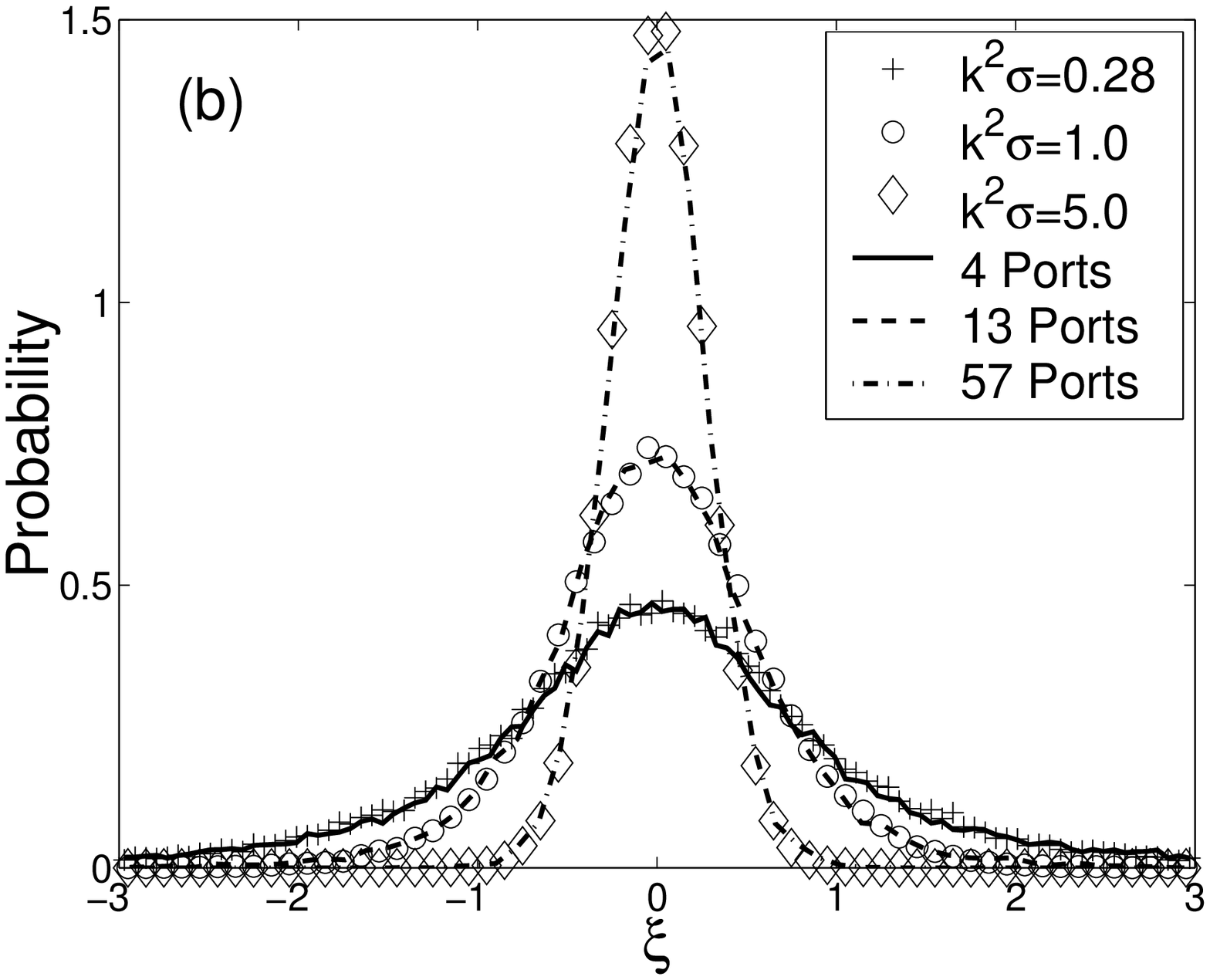}
\caption{Comparison between the impedance obtained from the one-port
lossy case and the multiple lossless case. (a) for the real part of the
impedance; (b)for the imaginary part of the impedance.}
\label{fig:lossaschan}
\end{figure}

We now briefly discuss the multiport  case with $M>2$ and  with
mismatch ($|\rho_R|>0$). As long as the assumption that the
eigenfunctions ($\eta$) and the eigenvalues ($\theta$ or $\phi$)
are independent is still true, $\langle |S_{11}|^2 \rangle$ is
related to the mismatch only through $\langle
\cos(\phi_k-\phi_l)\rangle$, similar to the expression in
Eq.~(\ref{eq:formulates11_goe}). The same series of steps specified
in the Appendix can be carried out to show that $\langle
\cos(\phi_k-\phi_l)\rangle$, as well as $\langle |S_{11}|^2
\rangle$, depend only on $|\rho_R|$ (and are independent of the
phase of $\rho _R$).  We have verified this
by numerical simulation using the impedance matrix generated from
(\ref{eq:xgue}) with up to seven channels.

\section{Summary}

We have generalized our random coupling model proposed in
Ref.~\cite{paperpart1} to
the multiport case. A similar impedance normalization is
applied to obtain the statistical properties of the multiport
chaotic scattering matrix.
The correlation
coefficients between eigenvalues are calculated explicitly and
agrees with the
random matrix theory.

We further incorporate the coupling parameters $\gamma_R$ and
$\gamma_X$ into the formulation of multiport scattering matrices and
present the formula for the averaged reflection coefficients
versus different values of coupling strength. We find that $|\rho_R|$,
which is a function of the two parameters above, characterizes the
transport process. For different pairs of $(\gamma_R$, $\gamma_X$), as
long as they yield the same value of $|\rho_R|$, the corresponding
averaged reflection coefficients are the same. This observation may
offer  a useful criteria for cavity design.

Using HFSS, we test the conclusions above using impedance data
calculated from direct numerical solution of Maxwell
Equations. The agreement between
the numerical results and the theoretical predictions convinces us that
our approach of impedance normalization successfully
recovers the
statistical ensemble for chaotic scattering in the multiple port
case.

\begin{acknowledgments}
We thank useful discussion with R. E. Prange, S. Fishman, J. Rogers and
S. Anlage as well as comments from Y. Fyodorov and P. Brouwer.
This work was supported in part by the DOD
MURI for the study of microwave effects under AFOSR Grant
F496200110374.
\end{acknowledgments}
\appendix*

\section{Evaluation of $\langle |S_{11}|^2\rangle $}

In this appendix, we will start from the one-port case, and obtain
an expression for  the phase of
 $S$ in term of the reflection coefficient $\rho _R$ defined in
Eq.~(\ref{eq:rhor}). Then, using Eq.~(\ref{eq:jointpdf}), we can
evaluate $\langle \cos(\phi_1-\phi_2) \rangle$ for the two-port in
the TRS and TRSB cases.

In the one-port case, $S$ can be expressed as
\begin{equation}
\begin{aligned}
S=\me^{\mj\phi} &= \frac{Z-Z_0}{Z+Z_0}\\
&=\frac{\mj(\gamma_X+\tilde \xi \gamma_R)-1}{\mj(\gamma_X+\tilde
\xi \gamma_R)+1},
\end{aligned}
\label{eq:s1}
\end{equation}
where $\tilde \xi$ is a zero mean, unit width, Lorentzian random
variable, which can be written as,
\begin{equation}
\tilde \xi=\tan \theta
\label{eq:xi1}
\end{equation}
with $\theta $ uniformly distributed in $[-\pi /2, \ \pi /2]$.
Putting Eq.~(\ref{eq:xi1}) into Eq.~(\ref{eq:s1}), we get
\begin{equation}
\me^{\mj\phi}=
\frac{(\gamma_R+\mj\gamma_X-1)\me^{\mj\theta}-(\gamma_R-\mj\gamma_X+1)
\me^{-\mj\theta}}
{(\gamma_R+\mj\gamma_X+1)\me^{\mj\theta}-(\gamma_R-j\gamma_X-1)
\me^{-\mj\theta}}. \label{eq:s2}
\end{equation}
Introducing $\rho_R$ such that
\begin{equation}
\gamma_R+\mj\gamma_X-1 =\rho_R(\gamma_R+\mj\gamma_X+1),
\end{equation}
and defining
\begin{equation}
\me^{-\mj\alpha}=\frac{\gamma_R-\mj\gamma_X+1}{\gamma_R+\mj\gamma_X+1},
\end{equation}
we obtain a compact expression for $\phi$ in term of $\theta$ and
$\rho_R$,
\begin{equation}
\begin{aligned}
\me^{\mj\phi}
&=\frac{\rho_R-\me^{-\mj(2\theta+\alpha)}}{1-\rho_R^*
\me^{-\mj(2\theta+\alpha)}}
\\
&=\me^{\mj\phi_{R}} \me^{-\mj2\theta '}
\frac{1+|\rho_R|\me^{\mj2\theta '}}{1+|\rho_R|\me^{-\mj2\theta
'}},
\end{aligned}
\label{eq:s3}
\end{equation}
where $2\theta '=(2\theta+\alpha+\pi+\phi_R)$. Since $\alpha$ and
$\phi_R$ depend only on the coupling coefficient $\gamma_R$ and
$\gamma_X$, and $2\theta$ is uniformly distributed in $[0,\ 2\pi
]$, the angle $2\theta '$ is also uniform in $[0,2\pi ]$. Thus,
\begin{equation}
\begin{aligned}
P_{\phi}(\phi) &=P_{2\theta'}(2\theta')|\frac{d (2\theta')}{d \phi}|\\
&= \frac{1}{2\pi}
\frac{1}{1+|\rho_R|^2-2|\rho_R|\cos(\phi-\phi_R)}.
\end{aligned}
\end{equation}

The relation between $\phi$ and $2\theta '$ also holds true for
multi-port cases. Furthermore, from the joint probability density
function of $2\theta_1$ and $2\theta_2$ in
Eq.~(\ref{eq:jointpdf}), which is only a function of the
difference of two angles, we find that $2\theta_1'$ and
$2\theta_2'$ have the same joint distribution specified in
Eq.~(\ref{eq:jointpdf}). Thus we can evaluate
\begin{equation}
\begin{aligned}
\langle \cos(\phi_1-\phi_2) \rangle &=Re[\me^{\mj\phi_1-\mj\phi_2}]\\
&=Re[\frac{\me^{-\mj 2\theta_1'}+|\rho_R|}{1+|\rho_R|\me^{-\mj
2\theta_1'}} \frac{\me^{\mj
2\theta_2'}+|\rho_R|}{1+|\rho_R|\me^{\mj 2\theta_2'}}],
\end{aligned}
\label{eq:cosave1}
\end{equation}
by using the joint distribution of $2\theta_1'$ and $2\theta_2'$,
$P_{\beta}(2\theta_1,2\theta_2) \propto |\me^{\mj
2\theta_1'}-\me^{\mj 2\theta_2'}|^{\beta}$, where $\beta=1$
corresponds to the TRS case, and $\beta=2$ for TRSB case.

Introducing $\psi_1=2\theta_1'$, $\psi_2=2\theta_2'$, and their
difference $\psi_-=\psi_1-\psi_2$, we obtain for the average of
$\cos(\phi_1-\phi_2)$,
\begin{equation}
\begin{aligned}
\langle \cos(\phi_1-\phi_2)\rangle &= \iint \frac{\dif \psi_1
d\psi_2}{(2\pi)^2} P(\psi_1,\psi_2)\\
& Re[\frac{\me^{-\mj\psi_1}+|\rho_R|}{1+|\rho_R|e^{-\mj\psi_1}}
\frac{\me^{\mj\psi_2}+|\rho_R|}{1+|\rho_R|\me^{\mj\psi_2}}] \\
&=\int \frac{\dif \psi_-}{2\pi} P(\psi_-)\\
& Re[\int_0^{2\pi}
\frac{\psi_2}{2\pi}
\frac{\me^{-\mj(\psi_-+\psi_2)}+|\rho_R|}{1+|\rho_R|
\me^{-\mj(\psi_-+\psi_2)}}\\
&\frac{\me^{j\psi_2}+|\rho_R|}{1+|\rho_R|\me^{\mj\psi_2}}].
\end{aligned}
\label{eq:cosave2}
\end{equation}
The inner integral can be calculated by introducing a complex
variable $z=\me^{j\psi_2}$ in terms of which the inner integral
becomes
\begin{equation}
\begin{aligned}
\frac{1}{2\pi \mj} \oint \limits_{unit circle} \frac{\dif z
f(z)}{z (z+|\rho_R|\me^{-\mj\psi_-})},
\end{aligned}
\label{eq:innerintegral}
\end{equation}
where
$f(z)=(|\rho_R|z+\me^{-\mj\psi_-})(z+|\rho_R|)/(1+z|\rho_R|)$.
Evaluating this integral via the residues at the two poles within
the unit circle, $z=0$ and $z=-|\rho_R|\me^{-\mj\psi_-}$,
we obtain
\begin{equation}
\begin{aligned}
\langle \cos(\phi_1-\phi_2) \rangle &=\int_0^{2\pi} \frac{\dif
\psi_-}{2\pi} P(\psi_-)
\\
&[1-\frac{(1-|\rho_R|^4)(1-\cos\psi_-)}
{1+|\rho_R|^4-2|\rho_R|^2\cos\psi_-}].
\end{aligned}
 \label{eq:cosave3}
\end{equation}
For the TRS case, $P_{\psi_-}(\psi_-)=\pi|\sin({\psi_-}/{2})|/2$,
and Eq.~(\ref{eq:cosave3}) yields
\begin{equation}
\begin{aligned}
\langle \cos(\phi_1-\phi_2) \rangle&=
\frac{|\rho_R|^4+2|\rho_R|^2-1}{2|\rho_R|^2}
\\
&+\frac{(1-|\rho_R|^2)^3}{4|\rho_R|^3}\ln
\frac{1+|\rho_R|}{1-|\rho_R|}.
\end{aligned}
\label{eq:finalgoe}
\end{equation}
For the TRSB case,
$P_{\psi_-}(\psi_-)=2\sin^2(\psi_-/2)=(1-\cos\psi_-)$, and
(\ref{eq:cosave3}) yields
\begin{equation}
\langle \cos(\phi_1-\phi_2)
\rangle=1-\frac{(|\rho_R|^2-1)(|\rho_R|^2-3)}{2}.
\label{eq:finalgue2}
\end{equation}

\end{document}